\documentclass[aps,pra,11pt,nofootinbib,showpacs,superscriptaddress]{revtex4}

\usepackage[english]{babel}
\usepackage{graphicx}
\usepackage{amsmath,bbm,amssymb,multirow,times,bm,color,amsthm}

 \theoremstyle{plain}
\newtheorem{oracle}{Data Input}

\newcommand{\bra}[1]{\langle{#1}|}
\newcommand{\ket}[1]{|{#1}\rangle}

\newcommand{\Ord}[1]{\mathcal{O}\left( #1 \right)}
\newcommand{\tOrd}[1]{\tilde{\mathcal{O}}\left( #1 \right)}

\definecolor{Pr}{rgb}{0.4,0.3,0.9}

\begin{document}

\title{Quantum computational finance:  quantum algorithm for  portfolio optimization }
\author{Patrick Rebentrost}
\email{cqtfpr@nus.edu.sg,pr@patrickre.com}
\affiliation{Centre for Quantum Technologies, National University of Singapore, 3 Science Drive 2, Singapore 117543}
\affiliation{Research Laboratory of Electronics,
Massachusetts Institute of Technology, Cambridge, MA 02139}
\author{Seth Lloyd}
\email{slloyd@mit.edu}
\affiliation{Department of Mechanical Engineering, Massachusetts Institute of Technology, Cambridge, MA 02139}
\affiliation{Research Laboratory of Electronics,
Massachusetts Institute of Technology, Cambridge, MA 02139}

\begin{abstract}
We present a quantum algorithm for portfolio optimization.  We discuss the market data input, the processing of such data via quantum operations, and the output of financially relevant results. Given quantum access to the historical record of returns, the algorithm determines the optimal risk-return tradeoff curve and allows one to sample from the optimal portfolio. The algorithm can in principle attain a run time of ${\rm poly}(\log(N))$, where $N$ is the size of the historical return dataset. Direct classical algorithms for determining the risk-return curve and other properties of the optimal portfolio take time ${\rm poly}(N)$ and we discuss potential quantum speedups in light of the recent works on efficient classical sampling approaches.  
\end{abstract}

\maketitle

\section{Introduction}

Quantum computing promises advantages for certain problems, including prime factoring, hidden subgroup problems, and unstructured search \cite{Nielsen2000}. Quantum computers perform favorably the task of processing vectors in large-dimensional Hilbert spaces. For $n$ quantum bits (qubits), 
the corresponding Hilbert space is $2^n$ dimensional and a normalized quantum state over $n$ qubits is fully described by $2^n-1$ 
complex numbers. Many basic quantum algorithms, such as the quantum Fourier transform, operate 
with run time $\Ord{{\rm poly}(n) }$ while manipulating all the coefficients classically would take $\Ord{2^n}$ 
operations. Quantum computing also provides speedups for 
sparse matrix inversion via the Harrow-Hassidim-Lloyd (HHL) algorithm \cite{Harrow2009,Childs2017linear}, 
an algorithm that finds application, e.g., in
data fitting \cite{Wiebe2012} and machine learning \cite{Rebentrost2014}.  Because
algorithms such as the quantum matrix inversion algorithm provide the solution in the form of a quantum state, care must be taken to insure that the algorithm can reveal the desired output in time $\Ord{{\rm poly}(n) }$ \cite{aaronson2015read}.  

Modern finance employs large amounts of computational resources for 
a variety of tasks. Computers are used for example for the analysis of historical data, high-frequency 
trading, pricing of exotic financial derivatives, portfolio optimization and risk management 
\cite{Glasserman2003,Follmer2004,Hull2012,Green2015}. The complexity often arises from the analysis of time series of asset prices over long periods
involving many assets, motivating the use of quantum computing for financial problems.
Quantum mechanics in combination with finance has been considered before. One avenue is the application of 
techniques from quantum physics, such as path integrals, to options pricing \cite{Baaquie2004,Haven2002}. With the advent of intermediate-scale quantum computers,  employing their power in finance is becoming more and more viable. 
Recently, the quantum amplitude estimation algorithm was shown to allow up to quadratic speedups for derivative pricing \cite{Rebentrost2018finance} and risk management \cite{Woerner2018}.
Several financial optimization problems have been considered by 1Qbit \cite{rosenberg2016finding} 
in the context of the D-Wave machine and the quantum annealing paradigm. In the same context, portfolio
optimization was  discussed in \cite{Marzec2016,Venturelli2018}.
To this point, no gate model quantum algorithm for portfolio optimization has been exhibited.

Here, we present quantum algorithms for portfolio management, specifically 
portfolio optimization \cite{Markowitz1952} and risk analysis. 
Portfolio optimization can be phrased as an equality-constraint quadratic programming problem:
one finds the portfolio that maximizes expected return for fixed standard deviation (risk), or, equivalently, that minimizes the standard deviation of the return for fixed expected return. 
The graph of maximal returns for fixed risk/minimum risks for fixed returns is called the
risk-return curve, see \cite{EfficientFronierWiki} for a figure.    Determining the risk-return curve and identifying the
portfolio that maximizes return for a given risk is a central task of classical portfolio
management theory.  Quadratic optimization is a problem that is known to be susceptible to speedup via quantum computation: here, we apply the quantum
linear systems algorithm and its variations \cite{Harrow2009,Childs2017linear} to obtain the the risk-return curve and
to identify the minimal-risk portfolio for given desired return.  The optimal portfolio is presented
as a quantum state which can then be sampled to give a sparse portfolio that is close
to optimal.  

The quantum portfolio optimization algorithm can exhibit a run time that is logarithmic in the number of assets and the number of time steps.  Suppose that there are $N$ possible investments and $T$ time instances, and our data set consists of $T$ $N$-dimensional vectors of the historic returns for those investments
stored in random access memory (RAM).    
Direct classical algorithms for portfolio optimization proceed
by constructing the $N\times N$ covariance matrix for the returns, using $\Ord{TN^2}$ operations/calls to RAM, and
implementing its pseudo-inverse ($\Ord{N^2}$ operations/memory calls) to construct the optimal
risk-return curve and to find the portfolio that maximizes return for a fixed value of risk.   
The quantum algorithms for portfolio management presented here can find the risk-return curve
and reveal the quantum state corresponding to the optimal portfolio in time $\Ord{ {\rm poly} \log(TN)}$.
If the relevant data is stored in quantum random access memory (qRAM) \cite{Lloyd2013,Giovannetti2008,Giovannetti2008_2,Martini2009}, so
that the quantum computer can access it in quantum parallel, then a quantum computer
can implement the HHL linear systems algorithm and its variants \cite{Harrow2009,Childs2017linear}, quantum walk Hamiltonian simulation methods, and the quantum state exponentiation method \cite{Lloyd2014,kimmel2017hamiltonian} to implement the matrix pseudo-inverse and to solve the quadratic optimization problem.  The quantum
algorithm for performing the pseudoinverse can take $\Ord{{\rm poly} \log(TN)}$ operations/memory calls.
Of course, loading the data set
into the qRAM to begin with takes time $\Ord{TN}$: the logarithmic run time
refers to the analysis of the data once it has been loaded into memory. The output of the quantum
algorithm is the optimal risk-return curve, together with quantum representations
of the optimal portfolios that attain the maximum return for each value of the risk.
The algorithm returns the optimal portfolio as a quantum state vector (dimension $N$).
To determine the exact composition of the optimal portfolio would take time $\Ord{N}$ to determine each component of that state vector.    We can, however, sample from the optimal
portfolio: the probability of returning a particular stock is proportional to (the square of) its weight in the portfolio.   By standard results on Monte Carlo sampling, a portfolio constructed
from a sample of $M$ elements taken from the optimal portfolio has an expected standard
deviation within $\Ord{1/\sqrt M}$ of the minimum standard deviation.  Moreover, we can make
quantum measurements on the optimal portfolio state to determine its weight in different sectors, e.g., energy, telecommunications, technology,
etc.
We can also compare the expected return and variance of any specific portfolio with that of the optimal one, 
to determine the degree of sub-optimality.
Finally, we can map out the efficient frontier, that is the optimal risk-return trade-off curve, 
in a time $\Ord{\log TN}$, where 
$T$ is the time horizon (number of time steps) and $N$ is the number of assets in the portfolio.
We discuss methods for loading the financial data via quantum random access memory and the preparation of the necessary input data structures, which are the expected return vector and the covariance matrix.   
Finally, we discuss the quantum
algorithms given here in the context of recent quantum-inspired classical algorithms
that also give a logarithmic run time, for example for recommendation systems, principal component analysis, and supervised clustering \cite{Kerenidis2017,Tang2018,Tang2018_2}.

\section{Market data}

\subsection{Prices and returns}
The sheer size of the financial market means that a vast amount of data is processed by the 
electronic exchanges. For each asset, such as stocks, bonds, options, and so on, the prices are 
moving on ever-shortening time scales,  as short as milliseconds or below. 
For managing a diversified portfolio or for performing a quantitative analysis of the 
economy \cite{Bai2003}, a large number of  high-dimensional vectors have to be processed.
Each high-dimensional vector represents the time series of a particular asset. 
Concretely, we have $N$ asset prices $\Pi_s(t)$ over a discretized time interval $T'\in \mathbbm Z_+$, taken to be a positive integer, where $s=[N]$ and $t=[T']$ and by definition $[x]:=1,\dots,x$. Here,
the time difference is normalized to $1$ which depending on the desired application can be below milliseconds.

In addition, a ``return" of a financial asset is a percentage profit or loss over a defined time period.
Let this time period be a positive integer $\Delta t \in \mathbbm Z_+$.
The return for an asset $s$ at time $t$ is here defined as
\begin{equation}
y_s(t):=\frac{\Pi_s(t) - \Pi_s(t-\Delta t)}{\Pi_s(t-\Delta t)}.
\end{equation}
The time label for the return is $t=\Delta t+1, \dots, T'$, as the prices $\Pi_s(t)$ are by assumption only given for $[T']$.
We can redefine the time label for the returns as $t=[T]$, where $T=T'-\Delta t$.

\subsection{Expected return and covariance}
The important data structures in this work are the vector of expected returns and the covariance matrix. 
Arrange the historical returns for the different assets in a vector $\vec y(t)$. We denote the corresponding random variable of the  returns at a future time $\Delta t$ from today by $\vec Y$.
The expected return is defined as 
\begin{equation}
\vec R^{\rm id} := \mathbbm{E}\left [\vec Y\right].
\end{equation}
Based on the historical prices, we obtain an estimate of this expected return via 
\begin{equation}
\vec R = \frac{1}{T} \sum_{t= 1}^{T} \vec y(t).
\end{equation}
Assuming $\sigma_s$ is the variance of the returns of asset $s$, the standard deviation of the sampled $\vec R_s$ is  
$\Ord{\sigma_s/\sqrt{T}}$.
Similarly, the covariance matrix is defined via the mean-normalized returns as 
\begin{equation}\label{eqCov}
\Sigma^{\rm id} = \mathbbm{V} \left [ \left (\vec Y - \vec R^{\rm id}\right) \left (\vec Y- \vec R^{\rm id} \right)^T \right],
\end{equation}
and an estimate can be obtained from the samples as
\begin{equation}\label{eqSampleCov}
\Sigma = \frac{1}{T-1} \sum_{t=1}^{T} ( \vec y(t) -\vec R ) (\vec y(t) -\vec R )^T.
\end{equation}

\section{Quantum market data}

\subsection{Quantum data access}
\label{secQRAM}

For the data input into the quantum computation, we follow closely the discussion in \cite{Lloyd2013,Giovannetti2008,Giovannetti2008_2,Martini2009}.
Let $\tilde d_i$ be given data with $i=[N_{\rm qRAM}]$.  Take $d_i$ to be the $m$-bit approximation for these data and assume that the implied error is smaller than other errors in the problem and the quantum algorithm. We assume access to the following query operation
\begin{equation}\label{eqOracle}
\sum_{i=1}^{N_{\rm qRAM}} \alpha_{i} \ket{i} \ket{0^{\otimes m}} \to \sum_{i=1}^{N_{\rm qRAM}} \alpha_{i}  \ket i \ket{ d_i}, 
\end{equation}
where $ \alpha_{i} \in \mathbbm C$ are arbitrary amplitudes.
One way to implement such an operation is via quantum random access memory (qRAM) \cite{Giovannetti2008,Giovannetti2008_2}. Assume that the data are stored in a classical information storage medium such as a CD or classical 
RAM. The value $d_i$ of the data is stored in the $i$'th cell of the memory. Classical RAM can be turned into qRAM by a set of quantum switches allowing the 
memory cells to be accessed. The quantum switches are arranged as a branching tree of with 
depth $\Ord{\log N_{\rm qRAM}}$ and width $N_{\rm qRAM}$. In addition, we need a quantum register with $\log N_{\rm qRAM}$ qubits, 
whose contents label the memory cell to be accessed.
Quantum RAM allows a quantum-parallel access of the data. 
As shown in \cite{Giovannetti2008,Giovannetti2008_2}, qRAMs have 
the surprising feature that even though all $\Ord{N_{\rm qRAM}}$ switches in the array participate in a memory call in 
quantum parallel, the expected number of switches that are actually switched or activated is 
$\log N_{\rm qRAM}$: consequently, the energy requirements and amount of decoherence induced by 
a memory call is proportional to the logarithm of the memory size.
When we compare the performance of a quantum algorithm to a classical algorithm, we assume that 
data can be accessed via classical RAM or quantum RAM, respectively. 

There are various potential ways of  the financial data being presented as an input to the quantum computation. 

\subsection{Expected return and covariance}
\label{secPriceOracle}

We start with a setting where the prices are stored and accessed via quantum RAM.
\begin{oracle}[Prices]\label{oraclePrices} 
Assume the qRAM availability of the prices $\Pi_s(t)$ to $m$-bit approximation, where $s=[N]$ and $t=[T]$, and $m$ is such that the inaccuracies are negligible compared to other errors.
\end{oracle}
Based on this data input,  we show how to create the expected return and the covariance matrix for the quantum finance algorithms.
Input (\ref{oraclePrices}) provides an operation 
\begin{equation}
\ket s \ket t \ket{0^{\otimes m}} \to \ket s \ket t \ket{\Pi_s(t)}. 
\end{equation} 
From this operation the single-step returns can be computed and stored in a $m$-bit encoding by a reversible operation on the qubit registers
\begin{eqnarray}
\ket {\Pi_s(t)} \ket{\Pi_s( t-\Delta t)} \ket{0^{\otimes m}} &\to& \ket{\Pi_s(t)} \ket {\Pi_s( t-\Delta t)}  \ket {y_s(t) }.
\end{eqnarray} 
Here, $| y_s(t) \rangle$ is a multi-qubit state storing $y_s(t)$ in binary representation. 
Thus, with small polynomial overhead, we are able to construct the operation
\begin{eqnarray}
\ket s \ket t \ket{0^{\otimes m}} &\to&\ket s \ket t \ket{ y_s(t)},
\end{eqnarray} 
for $t=[T]$.
To obtain a quantum state related to the return vector $\vec y$, we perform a controlled rotation of an ancilla qubit as
\begin{equation}
| y_s(t) \rangle \ket 0 \to | y_s(t) \rangle \left ( \sqrt{1-\delta^2 y_s(t)^2} \ket 0 + \delta y_s(t) \ket 1 \right),
\end{equation}
where we choose $0<\delta$ such that $\vert y_s(t)\vert \delta \leq 1$ for all $s,t$.
We apply the sequence of price oracle call, return calculation, and controlled rotation to the superposition $\frac{1}{\sqrt{TN}} \sum_{t=1}^T \sum_{s=1}^N |t\rangle  |s\rangle$. Uncomputing the data registers and 
measuring the ancilla in state $|1\rangle$ in such a superposition then gives the state 
\begin{eqnarray}
\frac{1}{\vert y \vert} \sum_{t=1}^T \sum_{s=1}^N  y_s(t) \ket t \ket s
= \frac{1}{\vert y \vert} \sum_{t=1}^T  \vert \vec y(t) \vert \ket t \ket{\vec y(t)}
=: |\chi \rangle,
\end{eqnarray}
with $\vert y \vert^2 = \sum_{t,s} y_s^2(t) = \sum_{t} |\vec y(t)|^2$ . Here, we obtain the quantum states encoding the returns for the times $t$ in amplitude encoding 
\begin{equation} \label{eqVecYt}
\ket{\vec y(t) } :=  \frac{1}{\vert \vec y(t) \vert} \sum_{s=1}^N   y_s(t)  |s\rangle.
\end{equation}
The success probability of the ancilla measurement is $P_\chi=\delta^2 \sum_{s,t}  y_s(t)^2 / TN$.
When we assume most of the returns are $y_s(t)=\Theta(1)$, then $P_\chi= \Omega(1)$.

We now turn to the preparation of the vector of expected returns. 
We generate the vector of expected returns by first applying a Hadamard operation to all qubits of the time register of $\ket \chi$ and then measuring the computational zero state
\begin{eqnarray}\label{eqAverageReturn}
\frac{1}{\sqrt{T}} \sum_{t=1}^{T} \langle t| \chi \rangle &\to&
\frac{1}{\vert y' \vert } \sum_{s=1}^N \left (\sum_{t=1}^T y_s(t) \right ) |s\rangle 
\nonumber \\
&\equiv&  \ket{\vec R }.
\end{eqnarray}
Here, $\vert y' \vert^2 = \sum_s (\sum_t y_s(t))^2$. The success probability of this measurement is given by 
$P_R = \vert y'\vert^2/(T\vert y \vert^2$). Again, if $y_s(t)=\Theta(1)$, then the success probablity $P_R= \Omega(1)$.
A simple quantum finance algorithm can be obtained by sampling of the quantum state
$|\vec R\rangle$. A particular stock $|s\rangle$ is measured with the probability 
$\left ( \frac{\sum_{t=1}^Ty_s(t)}{\vert y'\vert} \right )^2$. Thus, sampling can give an indication of the assets with the largest returns, especially if certain assets dominate the return vector.

In addition, we can prepare a quantum density matrix proportional to the covariance matrix of the returns. 
In Appendix \ref{appendixCov}, we show how to prepare a state $\ket{\tilde \chi} = 
\frac{1 }{\vert \tilde y \vert} \sum_{t,s} \ket t \ket s \left ( y_s(t) - \frac{1}{{T}} \sum_{t'}   y_s(t') \right )$, with $\vert \tilde y \vert^2 = \sum_{t,s}  \left ( y_s(t) - \frac{1}{{T}} \sum_{t'}   y_s(t') \right )^2 \equiv (T-1) {\rm tr} \Sigma$,
that amplitude-encodes the mean-normalized returns, in contrast to $\ket \chi$ above. If we take a partial trace over the time register of this state $\ket {\tilde \chi}$, we obtain the (normalized) covariance matrix in the subspace of the asset label 
\begin{eqnarray}\label{eqQuantumCovariance}
{\rm Tr}_1 \{ \ket{\tilde \chi} \bra{ \tilde \chi} \} &=& 
\frac{1}{\vert \tilde y \vert^2}  \sum_{s,s'=1}^N  \sum_{t=1}^T \left (y_s(t)- \frac{1}{T} \sum_{t'} y_s(t')\right) \left(y_{s'}(t)- \frac{1}{T}\sum_{t'} y_s(t')\right )  |s\rangle  \langle s'| \nonumber \\
&=& \frac{\Sigma}{{\rm tr} \Sigma}  .
\end{eqnarray}
The average return Eq.~(\ref{eqAverageReturn}) and the covariance matrix Eq.~(\ref{eqQuantumCovariance}) are the main quantities 
to be used for the portfolio optimization. 
Another simple quantum algorithm follows. By measuring this covariance matrix one can obtain samples of the assets with the largest variances and pairs of assets with the largest covariances. 

\subsection{Enhanced data input for prices and expected returns}
\label{secKPData}

For the quantum portfolio optimization algorithm, 
it will be advantageous to assume a particular data structure embedded in qRAM. 
In \cite{Kerenidis2017}, Kerenidis and Prakash (KP) use a specific binary-tree data structure to store vectors.
Each node of the tree stores the subnorms, i.e., the sums of squares of the entries of the vector corresponding to the nodes below. 
The lowest nodes store the squares of the vector elements and their signs. 
A similar idea of storing subnorms in quantum RAM was previously discussed in \cite{Lloyd2013}. 
Such a natural data structure can be embedded in 
quantum RAM itself and has the advantage that it can be updated efficiently and continuously and
allows for a deterministic state preparation analogous to Grover-Rudolph \cite{Grover2002} including signs.

\begin{oracle} [KP for prices]\label{oracleKPPrices}
Assume a KP data structure embedded in qRAM for present day's asset prices $ \Pi_s$, $s=[N]$.
\end{oracle}
This data structure allows the preparation without measurement of the state
\begin{equation}
\ket{0^{\otimes  \lceil \log N \rceil}} \to \frac{1}{\vert \Pi \vert}\sum_{s=1}^N \Pi_s \ket s =: \ket{ \Pi}.
\end{equation}
with $\vert \Pi \vert^2  = \sum_{s=1}^N \Pi_s^2$. 
 The next data input assumption is stronger, since we assume that the average returns are already precomputed and stored.
 \begin{oracle}[KP for average returns] \label{oracleKPReturn}
 Assume a KP data structure embedded in qRAM for the average returns $R_s$, $s=[N]$.
\end{oracle}
Using this data input, we can perform deterministically the operation 
$\ket{0^{\otimes \lceil \log N \rceil}} \to \ket{\vec R }$. No measurement is required in contrast to Eq.~(\ref{eqAverageReturn}).

\subsection{Precomputed covariance matrix}

In Sec.~\ref{secPriceOracle}, we prepared the covariance matrix as a density matrix. Alternatively, in some cases the covariance matrix is already precomputed and stored. Assuming the corresponding quantum access to the matrix elements allows the use of well-studied quantum walk methods for employing $\Sigma$ in a quantum algorithm. 
\begin{oracle}[Covariance matrix] \label{oracleCov}
Assume access to the elements $\Sigma_{jk}$ by querying an oracle as follows:
\begin{eqnarray}
\ket{jk}\ket{0^{\otimes m}}\rightarrow \ket{jk}\ket{\Sigma_{jk}}.
\end{eqnarray}
\end{oracle}
In addition, for sparse matrices, we assume the following data input. 
\begin{oracle} [Sparsity] \label{oracleSparse} 
Let $i=[N]$ and $l=[s_\Sigma]$, where $s_\Sigma$ is the sparsity of $\Sigma$.
Assume access to the operation  $ \ket{i,l} \to \ket{i, g_\Sigma(i,l)}$, where the efficiently computable function $g_\Sigma(i,l)$ gives the column index of the $l$-th nonzero element of row $i$ of matrix $\Sigma$.
\end{oracle}
The use of these operations will be discussed in Section \ref{sectionSimCov}.

\label{secKPData}

\section{Portfolio optimization}

H. Markowitz is recognized for introducing the modern version of portfolio theory \cite{Markowitz1952}. 
An optimal investment strategy achieves a certain desired return while risk is minimized, which is the setting we consider here.  Equivalently, the problem can be posed as attaining a certain desired risk while maximizing the return. 
This simple idea of risk-return optimization leads to the notion of portfolio 
diversification, that is, the optimal portfolio is likely one investing in many relatively uncorrelated assets. 
In contrast, when the chosen strategy is to only optimize returns, one obtains that the optimal portfolio 
selects only the single asset (or a few assets) with the highest expected return, irrespective of the risk involved. 

The assets are specified by a vector of today's asset prices given by $\vec \Pi$, the historical expected return vector of the assets $\vec R$, and the historical correlations in the assets given by $\Sigma$.
Note the sampling errors when using $\vec R$ versus $\vec R^{\rm id}$ and $\Sigma$ versus $\Sigma^{\rm id}$, which carry over to the quantum algorithm. 
Let the total wealth of an investor be given by $\xi$.
This wealth shall be allocated in the assets, specified by a portfolio allocation vector $\vec w$
which gives the amount invested in each asset. 
Here, we allow for short-selling, which means that the entries of $\vec w$ can be negative. 
Based on the total wealth, we obtain the constraint for the portfolio given by $\xi = \vec \Pi^T \vec w$.
 Based on the historical returns, the portfolio has an expected return given by $\vec R^T \vec w$. 
In addition, the portfolio risk can by specified as $\vec w^T \Sigma \vec w$. The portfolio optimization problem considered here is that the investor would like to achieve an expected return 
 $\mu = \vec R^T \vec w$, while minimizing the portfolio risk $\vec w^T \Sigma \vec w$. 
Thus, the problem is an equality-constrained quadratic program as follows:
\begin{eqnarray}
&\min_{\vec w}& \vec w^T \Sigma \vec w \\
&{\rm s.t. }& \vec R^T \vec w = \mu, \label{eqEqualityConstraintReturn} \\ 
&&\vec \Pi^T \vec w =\xi. \label{eqEqualityConstraintWealth}
\end{eqnarray}
We may also redefine the portfolio vector such that the last equality becomes
$\vec 1^T \vec w =\xi$, where $\vec 1$ is the vector of all $1$'s.
Introducing the Lagrange multipliers $\theta$ and $\eta$ 
for the equality constraints Eqs.~(\ref{eqEqualityConstraintReturn}) and 
(\ref{eqEqualityConstraintWealth}),
the linear equation system to solve for the optimization problem is given by $M \vec x = \vec b$, which is defined as
\begin{equation}\label{eqPortfolioOptimizationLinearEquation}
\left ( \begin{array}{ccc}
0 & 0 & \vec R^T \\
0 & 0 & \vec \Pi^T \\
\vec{R} &\vec \Pi  & \Sigma
\end{array}
\right ) 
\left ( \begin{array}{c}
\eta \\
\theta \\
\vec w
\end{array}
\right )
=
\left ( \begin{array}{c}
 \mu \\
 \xi \\
\vec 0
\end{array}
\right ).
\end{equation}
Quantum mechanically, we solve the corresponding linear system
$\hat M \ket{ \eta, \theta, \vec w}= \ket{\mu, \xi,\vec 0} $ via
the HHL algorithm and its variants \cite{Harrow2009,Childs2017linear}.
The eigenvalues of the matrix $\hat M=M/{\rm Tr}M$ are denoted by $\lambda_j$ and 
the eigenvectors by $|u_j\rangle$. The right-hand side of 
Eq. (\ref{eqPortfolioOptimizationLinearEquation}) becomes the normalized 
quantum state $\ket{\mu, \xi,\vec 0} $.
Such a state is easy to prepare since it only consists of the two quantities $\mu$ and $\xi$.
The solution the HHL algorithm provides is the normalized quantum state:
\begin{equation}
\ket{ \eta, \theta, \vec w} = \frac{1}{ \vert v\vert } \sum_{j:\lambda_j \geq 1/\kappa} \frac{\beta_j}{\lambda_j} |u_j\rangle,
\end{equation}
with the norm $ \vert v\vert = \sqrt{\sum_{j:\lambda_j \geq 1/\kappa}( \beta_j / \lambda_j)^2}$
and $\beta_j := \langle u_j\ket{\mu, \xi,\vec 0} $.
Here, $\kappa$ is chosen to be a constant. 
The HHL algorithm projects onto the well-conditioned subspace with eigenvalues greater than 1$/\kappa$. 
In this way, the algorithm finds the pseudoinverse of $\hat M$ in such a way that only eigenvalues $\lambda_j \geq 1/\kappa$ are taken into account. Let us denote this pseudoinverse by $\hat M_\kappa^{-1}$. The procedure is equal to the actual pseudoinverse $\hat M^{-1}$ whenever $1/\kappa$ is smaller than the smallest eigenvalue $\vert \lambda_{\min}\vert$ of $\hat M$. Otherwise, $\hat M_\kappa^{-1}\ket{ \vec 0, \mu, \xi }$ approximates $\hat M^{-1}\ket{ \vec 0, \mu, \xi }$ to an error
\begin{equation}
\epsilon_\kappa := \left \vert \hat M_\kappa^{-1}\ket{ \vec 0, \mu, \xi } - \hat M^{-1}\ket{ \vec 0, \mu, \xi }\right \vert_2.
\end{equation}
Thus, efficient quantum portfolio optimization requires $\kappa = {O}\left({\rm poly}\left(\log{d}\right)\right)$, but also the requirement that $\hat M$ to be such that either (1) $\left|\lambda_{\min}\right| \geq 1/\kappa$, with no additional errors in finding the pseudoinverse, or (2) $\left|\lambda_{\min}\right| < 1/\kappa$ but with $\epsilon_\kappa = \Ord{\epsilon}$ so that the error $\epsilon_\kappa $ accumulates in accordance with an overall desired error $\Ord{\epsilon}$.

The success probability of preparing $\ket{ \eta, \theta, \vec w}$ is 
\begin{equation}
P_w = C^2 \sum_{j:\lambda_j \geq 1/\kappa} \left \vert \frac{\beta_j}{\lambda_j} \right \vert^2, 
\end{equation}
with a user-specified $C = \Ord{1/\kappa}$. 
We can determine $P_w$ itself from multiple runs of the linear systems algorithm.
To relate the quantum state  $ \ket{ \eta, \theta, \vec w}$ to the actual solution of (\ref{eqPortfolioOptimizationLinearEquation}), we multiply $\ket{ \eta, \theta, \vec w}$ by a factor
\begin{equation}\label{eqFactor}
\sqrt{\frac{P_w (\mu^2 +\xi^2)}{C^2}}\ {\rm tr}{\Sigma}, 
\end{equation}
which involves only known quantities. The trace ${\rm tr}{\Sigma}$ is estimated via Appendix \ref{appendixCov}, Eq.~(\ref{eqPtrSigma}).
When measuring properties of the portfolio we always multiply the result by this factor 
to obtain the actual desired value.  
To obtain a quantum state $|\vec w\rangle$ we project the state 
$\ket{ \eta, \theta, \vec w}$ onto the desired part.

A central step in the HHL algorithm is the controlled matrix exponentiation of $\hat M$, i.e., the quantum simulation of the unitary $e^{-i \hat M \Delta t}$ for the use in phase estimation. 
The matrix  $M=H_{\Sigma} + H_{\vec R}+H_{\vec \Pi}$ consists of three parts, 
(i) the covariance matrix part $H_{\Sigma}$, (ii) the expected return vector part 
$H_{\vec R}$, and (iii) the price vector $H_{\vec \Pi}$. The simulation of these parts is discussed in the following subsections. 
The matrix sum can be then simulated via the Lie product formula a small time step $\Delta t$
such that $\exp(-i M t) \approx \exp(-iH_{\Sigma} \Delta t)\exp(-iH_{\vec R} \Delta t)\exp(-iH_{\vec \Pi} \Delta t)$ with error $\Ord{\Delta t^2}$.

\subsection{Simulation of the covariance matrix} 
\label{sectionSimCov}

The simulation of the covariance matrix depends on the data input model and market conditions. 
First, if we assume Input (\ref{oraclePrices}), we prepare the covariance matrix as a quantum density matrix, see Eq.~(\ref{eqQuantumCovariance}). With this density matrix, we can use quantum state exponentiation techniques  \cite{Lloyd2014,rebentrost2018quantum}, which require the covariance matrix to be low rank. 
Such low-rank situations frequently occur in the financial context. Often market conditions exists where a relatively small number of underlying \textit{factors} are driving the asset prices \cite{Bai2003}. Such factors could be for example the interest rate policies of the Federal Reserve or the geopolitical climate. Some macroeconomic factors drive predominantly certain subsectors of the stock market. For example, the automobile or aviation sectors depend on consumer behavior, or the energy sector depends on the availability of fossil fuels, and so forth.

The strategy presented in reference \cite{Lloyd2014}
can be used to enact the matrix exponential of the covariance matrix. Let $\sigma$ be a quantum state in an $N$-dimensional Hilbert space
and  $\rho =\Sigma /{\rm tr} \Sigma$ in another $N$-dimensional Hilbert space. The task is to perform $e^{-i \rho t} \sigma e^{i \rho t} $, i.e., to use $\rho$ as a Hamiltonian.
With the $N^2 \times N^2$ dimensional swap matrix $S$
we can generate
\begin{equation}\label{eqSimulationTrick}
{\rm tr}_1 \{ e^{-i S \Delta t} \rho \otimes \sigma e^{iS\Delta t} \} = \sigma - i\Delta t [\rho, \sigma] + \Ord{\Delta t^2},
\end{equation}
where the trace is over the $\rho$ Hilbert space. To simulate $e^{-i \rho t}$ for a total time $t$ and to accuracy $\epsilon$, one needs $\Ord{t^2/\epsilon}$ copies of $\rho$. The rank of the effectively simulated density matrix is given by $r=\Ord{ t^2}$ \cite{rebentrost2018quantum}. A controlled simulation required for phase estimation can be implemented \cite{Lloyd2014,kimmel2017hamiltonian}.

In addition, we can also efficiently simulate the covariance matrix in other regimes and different forms of data input. Let Inputs (\ref{oracleCov}) and (\ref{oracleSparse}) be given. The elements of the covariance matrix have been measured and stored and the effective covariance matrix is sparse. 
In some cases, there are market conditions which correspond to such high-rank, sparse covariance matrices. An example of this case is when disjoint small subsets of stocks are driven by independent factors each.
An example situation would be if pairs of stocks follow similar trajectories (one of the pairs for example being the two leading sugary beverage companies.) These situations correspond to effectively sparse, high-rank covariance matrices. 

In this case, we can apply quantum simulation techniques for sparse Hamiltonians \cite{Berry2012,Berry2015,Low2017}, which employ Inputs (\ref{oracleCov}) and (\ref{oracleSparse}) and quantum walks. These techniques exhibit a query complexity of $\Ord{\frac{(s_\Sigma \tau)^{1+o(1)}}{\epsilon^{o(1)}} }$, where $s_\Sigma$ is the sparsity of $\Sigma$, $\tau = \Vert \Sigma \Vert_{\max} t$ with $t$ the simulation time and $\Vert \cdot \Vert_{\max}$ the maximum element norm, $\epsilon$ the accuracy in operator norm, and $o(1)$ denotes a quantity strictly dominated by a constant. The gate complexity of these techniques scales 
as $\tOrd{\log N}$ in the dimension of the matrix.

\subsection{Simulation of the matrices for returns and prices}

The portfolio optimization involves the equality constraints, encoded 
in the $N+2$ dimensional symmetric matrices containing a row and column with the $N$ dimensional return vector and the 
vector of the prices. The simulation is relatively straightforward  \cite{Childs2010}. 
Both correspond to a ``star" graph, 
where one site is connected to $N$ of others with given weights.
For the return matrix, the center of the star graph is site $\ket 1$. The two nonzero eigenvalues of the embedded return matrix $ H_{\vec R}$
are $\lambda_{\pm}^{R}=\pm \sqrt{\sum_{s=1}^N R_s^2} $ and the corresponding
eigenstates are $| {\lambda_{\pm}^R}
\rangle=\frac{1}{\sqrt{2}} \left( |1\rangle + \frac{1}{\lambda_{\pm}^R}
\sum_{s=1}^N R_s |s+2\rangle \right)$. 
We note again that the matrix $ H_{\vec R}$ is $N+2$ dimensional.
To enact the matrix exponential, one transforms into the eigenbasis of the weighted star, 
exponentiates the eigenvalues and transforms back into the computational basis \cite{Childs2010}. 
The transformation into the eigenbasis is accomplished by first performing the transformation 
$|1\rangle \to 1/\sqrt{2} (\ket 1 + \ket 3) $, $|3\rangle \to 1/\sqrt{2}( \ket 1 - \ket 3)$, 
and the identity on the other states, corresponding to a single Hadamard operation between states $\ket 1$ and $\ket 3$. Then perform any 
transformation satisfying $\ket 1 \to \ket 1$ and $\ket 3 \to \ket{\vec R}$.
Thus we require the Input (\ref{oracleKPReturn}) which allows a deterministic procedure of preparing $\ket{\vec R}$. 
For the price matrix $H_{\vec \Pi}$, we proceed analogously, with the center of the star graph being $\ket 2$ instead of $\ket 1$. We require Input (\ref{oracleKPPrices}) for performing a deterministic preparation of $\ket{\Pi}$.
This step may be simplified by redefine the portfolio vector such that $\vec 1^T \vec w =\xi$, which implies that only the uniform superposition has to be prepared instead of $\ket{\Pi}$.

\section{What to do with the resulting quantum state}

Once we have prepared the optimal portfolio state $\ket w$, we would like to obtain classically relevant
information from it. However, we would like to avoid measuring the full state, which would take $\Ord{N}$ 
copies and would eliminate any possibility for a $\Ord{\log N}$ performance.
In the following, we describe different ways of potentially using the quantum portfolio state. 

First, on the optimal portfolio state we can measure the risk $\xi= \bra w \Sigma \ket w$ of the optimal portfolio. 
Given the density matrix Eq.~(\ref{eqQuantumCovariance}), we can perform a 
swap test \cite{Buhrman2001,Lloyd2013} between that density matrix and $\ket {w} \equiv \ket{ w(\mu, \xi)}$. This allows to map out the risk-return trade-off curve in the two-dimensional $(\mu,\xi)$ space. The accuracy is $\epsilon$ at a gate complexity of $\tOrd{\log N/\epsilon^2}$.
The swap test can also be used to compare the optimal portfolio obtained by the quantum algorithm against another portfolio vector prepared as a quantum state.
Consider portfolio state $\ket{\tilde w}$  offered by a counterparty, say a hedge fund. We would like to make a decision to invest into this portfolio. The distance from the optimal portfolio $|w\rangle$ can be computed again via a swap test with gate complexity as above to a given accuracy $\epsilon$. If the overlap between the portfolios is sufficiently high, one can consider 
the investment $\ket{\tilde w}$ to be sufficiently rational. 

Another simple way of using the portfolio quantum state is by measuring projectors. We can measure $\langle P_{\rm sector} |w \rangle$ with $|P_{\rm sector}\rangle$ being a quantum state that projects into a certain subset of assets. 
This allows to obtain the weight of the optimal portfolio in a certain sector of the economy.

We can also take a sampling approach with respect to the portfolio state $\ket w$.
Assume for the present discussion, we have Input (\ref{oracleKPReturn}) for the average returns $R_j$. 
In addition, assume that the portfolio state that is the output of the portfolio optimization is sparse, i.e., only $s_w=\Ord{\log N}$ elements are non-zero. 
Multiple samples, say $M = \Ord{\log N}$, effectively obtain another sparse portfolio  $\ket {w'}$.
One sample gives an index $j\in[N]$ with probability $\vert w_j\vert^2$. Multiple samples obtain the numbers $M_j$ of occurrences of index $j$.  
The best approximation of the actual portfolio position based on the obtained information is $\vert w_j' \vert := \sqrt{M_j/M}$, where we lose the sign of the entry. 
The sign can be recovered by the following reasoning, which we call \textit{long/short assumption}. If the corresponding average return $R_j$ is positive (negative) then also the position in the optimal portfolio is positive (negative). The optimal portfolio goes long on the stocks that have a positive average return and goes short on the stocks that have a negative average return. Since we have query access to $R_j$ including the signs we can find the sparse portfolio $w_j' = {\rm sgn} (R_j) \sqrt{M_j/M}$. The estimators $\vert w_j' \vert^2$ have expectation value $\vert w_j \vert^2$ under the long/short assumption and variance  $\sigma^2_j = \frac{\vert w_j\vert^2(1-\vert w_j\vert^2)}{M}$.

Now compare the return and risk of this sampled portfolio to the optimal portfolio. Define a random variable $Z$ to be $ R_j/w_j'$ depending on the sample outcome $j$, where we assume again we have query access to  $R_j$. The difference to Tang \cite{Tang2018} is that we do not have query access to $w_j$ but rather have only the estimate $w_j'$. The expectation value is
given by 
\begin{equation}
\mathbbm E [Z] = \sum_j \vert w_j\vert^2 \frac{R_j}{w_j'} = \sum_j  \vert w_j\vert^2 \vert R_j\vert  \sqrt{\frac{M} {M_j}}  \xrightarrow{M_j/M\rightarrow \vert w_j\vert^2} \sum_j \vert w_j\vert \vert R_j \vert.
\end{equation}
Thus, given many samples and the correctness of the long/short assumption, the expectation value converges to the desired expected return of the optimal portfolio. As we only have a finite number of samples, we perform an error analysis. Start with the second moment
\begin{equation}
\mathbbm E [Z^2] = \sum_j \vert w_j\vert^2 \frac{R_j^2}{w_j'^2} = \sum_j  \vert w_j\vert^2  R_j^2  \frac{M} {M_j} \xrightarrow{M_j/M\rightarrow \vert w_j\vert^2} \sum_j R_j^2.
\end{equation}
The variance converges to the one expected from infinite samples \cite{Tang2018}.
Using $\sideset{}{'} \sum_j$ to indicate that only $j$'s are summed over that have $\vert w_j\vert^2>0$, we can analyse the finite-sample behavior as
\begin{eqnarray}
\mathbbm E [Z^2] &=& \sum_j \vert w_j\vert^2  R_j^2  \left \vert \frac{M} {M_j} - \frac{1}{\vert w_j\vert^2} + \frac{1}{\vert w_j\vert^2} \right \vert \leq \sum_j R_j^2 + \sideset{}{'} \sum_j  \vert w_j\vert^2  R_j^2  \left \vert \frac{M} {M_j} - \frac{1}{\vert w_j\vert^2} \right \vert \\
&=& \sum_j R_j^2 + \Ord{\sideset{}{'} \sum_j   R_j^2 \frac{ \sigma_j}{ \vert w_j\vert^2}}.
\end{eqnarray}
Thus, we have an additional variance of the return proportional to $ \sideset{}{'}\sum_j  R_j^2  \sqrt{\frac{(1-\vert w_j\vert^2)}{M\vert w_j\vert^2}}$. This variance behaves as expected considering that $ \vert w_j\vert^2 = \Ord{1/s_w}$ because of the sparsity assumption. We have used that if $\vert x - x'\vert \leq \epsilon$ then $\vert \frac{1}{x} - \frac{1}{x'}\vert =\Ord{\frac{\epsilon}{x^2}}$.

Now consider the risk of the sampled portfolio. First, note that
the covariance matrix $\Sigma$ is positive semi-definite and the minimized risk $\bra w \Sigma \ket w$ is always smaller than the sampled one $\bra {w'} \Sigma \ket {w'}$.
In addition, we show the following simple bound. Let $\vert \ket w - \ket{w'}\vert \leq \epsilon_w$. Then, $\vert \bra w \Sigma \ket w - \bra {w'} \Sigma \ket {w'}\vert = \Ord{\epsilon_w}$.
This follows from
\begin{eqnarray}
\vert \bra w \Sigma \ket w - \bra {w'} \Sigma \ket {w'}\vert &=& \vert \bra w \Sigma \ket w - \bra w \Sigma \ket {w'} +  \bra w \Sigma \ket {w'} - \bra {w'} \Sigma \ket {w'}\vert \\
&\leq& \vert \bra w \Sigma \ket w - \bra w \Sigma \ket {w'} \vert +  \vert \bra w \Sigma \ket {w'} - \bra {w'} \Sigma \ket {w'}\vert \leq 2 \Vert \Sigma \Vert \epsilon_w.
\end{eqnarray}
Moreover, we know that the variance of $M_j/M$ is $\vert w_j\vert^2 (1-\vert w_j\vert^2)/M$, thus the $l_2$ norm of the difference $\vert \ket w - \ket{w'}\vert$ is bounded as $\epsilon_w =\Ord{\sqrt{ s_w \vert w_{\max} \vert^2 (1-\vert w_{\max}\vert^2)/M}}$. If the probabilities are evenly distributed, $ \vert w_j\vert^2 = \Ord{1/s_w}$, then $\epsilon_w = \Ord{1/\sqrt{M}}$. Thus, the risk of the sampled portfolio has two contributions: the original risk of the optimal portfolio and the additional risk from the finite sampling. The finite sampling risk goes as $\Ord{1/\sqrt{M}}$. 

\section{Discussion and conclusion}
Quantum matrix inversion schematically prepares a state $|x\rangle = A^{-1} |b\rangle$ on which measurements $\langle x | O | x\rangle$ can be performed. Measuring all the elements of $|x\rangle$ itself would remove the possibility of an exponential speedup. It is thus crucial to find good measurement operators $O$ which give results that are hard to obtain otherwise and at the same time that give new insights on the state $|x\rangle$. Measuring the risk of an optimal portfolio and the binary decision to invest in an offered portfolio provide classical information that relies on global properties of the state. The full set of interesting measurements on the portfolio vector is a topic for  further research.

In principle, the algorithms presented here  can be efficient in the sense that they have 
a run time proportional to the number of qubits involved in the computations, i.e., the run time is $\Ord{\log N}$. 
Hardware requirements are $\Ord{N}$ to store and access the data. 
The conditions for such a logarithmic run time is that the input can be prepared efficiently and the criteria of quantum matrix inversion are satisfied \cite{Harrow2009,Aaronson2015}.
Classical Monte Carlo methods can be efficient, often closing potential exponential 
separations between quantum algorithms and naive classical methods.
In particular, classical methods prove powerful when certain assumptions can be made on the preprocessing of the input data, especially for low-rank situations \cite{Frieze2004,Tang2018,Tang2018_2},
where the singular value decomposition can be constructed from a polylogarithmic number of samples.   Such classical methods allow the performance of matrix
completion algorithms \cite{Tang2018} and principal component analysis \cite{Tang2018_2}
 in polylogarithmic time, and could be applied to the low-rank matrix pseudo-inverse as well.
Exponential separations between a quantum algorithm and sampling algorithms have been 
shown for example for oracle promise problems involving the evaluation of inner products of vectors processed via Fourier transforms 
\cite{Aaronson2009}. Sparse matrix inversion also shows the potential for exponential speedups, evidenced indirectly via 
encoding a universal quantum computation into a matrix inversion \cite{Harrow2009}. 
Classical sampling is hard, i.e, $\Omega(N)$, for example for obtaining a low-rank matrix approximation when no prior information is given about the matrix \cite{Yossef2003}. 

While exponential speedups are the most compelling, quantum computing can provide polynomial speedups in rather  generic settings, for example for search \cite{Grover1996}, Monte Carlo \cite{Szegedy2004,Montanaro2015}, semi-definite programming \cite{Brandao2017,Apeldoorn2017}, and generalized convex optimization \cite{Apeldoorn2018,Chakrabarti2018}. 
The results of \cite{Tang2018,Tang2018_2} also allow for polynomial speedups for the discussed algorithms.  It is thus believable that such speedups can be obtained for financial applications such as derivative pricing \cite{Rebentrost2018finance} or the present work of portfolio optimization. The financial context is particularly sensitive to  speedups in terms of the potential rewards. Small improvements in speed for particular financial calculations can have a large impact in terms of financial reward and thus motivate further study of the intersection of quantum computing and finance. 

\begin{acknowledgements}
P.~R. acknowledges support from Singapore’s Ministry of Education and National Research Foundation. S.~L. was supported by ARO. 
\end{acknowledgements}

\appendix

\onecolumngrid

\section{Preparing the covariance matrix}
\label{appendixCov}

The covariance matrix is defined over the mean-adjusted returns. Given the oracle in Eq.~(\ref{eqOracle}), we show here how to construct a quantum state 
that contains the mean-adjusted returns in amplitude encoding. The notation omits ancilla qubits in the $\ket 0$ state. 
We proceed as
\begin{eqnarray}
\ket{0\dots 0} &\to& \frac{1}{\sqrt{2 TN}} \sum_{t,s} \ket t \ket s (\ket 0_a + \ket 1_a) \\
\text{(controlled Hadamards)} &\to&\frac{1}{\sqrt{2 TN}} \sum_{t,s} \ket t \ket s \left (\ket 0_a \ket 0_b + \frac{1}{\sqrt{T}} \ket 1_a \sum_{t'} \ket {t'}_b \right)\\
\text{(oracle query)} &\to&\frac{1}{\sqrt{2 TN}} \sum_{t,s} \ket t \ket s \left (\ket 0_a \ket 0_b \ket 0_c+ \frac{1}{\sqrt{T}} \ket 1_a \sum_{t'} \ket {t'}_b \ket {y_s(t')}_c \right)
\end{eqnarray}
\begin{eqnarray}
\text{(controlled rotation)} &\to&\frac{1}{\sqrt{2 TN}} \sum_{t,s} \ket t \ket s \left (\ket 0_a \ket 0_b \ket 0_c \ket 1_d+ \right . \\
&& \left . \frac{1}{\sqrt{T}} \ket 1_a \sum_{t'} \ket {t'}_b \ket {y_s(t')}_c \left( \sqrt{1-\delta^2 y_s(t')^2} \ket 0_d + \delta y_s(t') \ket 1_d  \right)\right)\\
\text{(uncompute)} &\to&\frac{1}{\sqrt{2 TN}} \sum_{t,s} \ket t \ket s \left (\ket 0_a \ket 0_b \ket 1_d+ \right . \\
&& \left . \frac{1}{\sqrt{T}} \ket 1_a \sum_{t'} \ket {t'}_b \left( \sqrt{1-\delta^2 y_s(t')^2} \ket 0_d + \delta y_s(t') \ket 1_d  \right)\right)
\end{eqnarray}
\begin{eqnarray}
\text{(controlled Hadamards)} &\to&\frac{1}{\sqrt{2 TN}} \sum_{t,s} \ket t \ket s \left (\ket 0_a \ket 0_b \ket 1_d+ \right . \\
&& \left . \frac{1}{{T}} \ket 1_a \sum_{t't''} (-1)^{t'\cdot t''}\ket {t''}_b \left( \sqrt{1-\delta^2 y_s(t')^2} \ket 0_d + \delta y_s(t') \ket 1_d  \right)\right).
\end{eqnarray}
Rotate another ancilla, with $0< \delta \leq \max_{s,t} \vert y_s(t)\vert$,
\begin{eqnarray}
&\to&\frac{1}{\sqrt{2 TN}} \sum_{t,s} \ket t \ket s \left (\ket 0_a \ket 0_b \ket 1_d \left( \sqrt{1-\delta^2 y_s(t)^2} \ket 0_e + \delta y_s(t) \ket 1_e  \right)+ \right . \\
&& \left . \frac{1}{{T}} \ket 1_a \sum_{t't''} (-1)^{t'\cdot t''}\ket {t''}_b \left( \sqrt{1-\delta^2 y_s(t')^2} \ket 0_d + \delta y_s(t') \ket 1_d  \right) \ket{1}_e \right).
\end{eqnarray}
Postselect on the state
\begin{equation}
\frac{1}{\sqrt 2}\left ( \ket 0_a + \ket 1_a \right) \ket 0_b \ket 1_d \ket 1_e.
\end{equation}
The postselection obtains
\begin{eqnarray}
&\to&\frac{1 }{ \vert \tilde y \vert} \sum_{t,s} \ket t \ket s \left ( y_s(t) - \frac{1}{{T}} \sum_{t'}   y_s(t') \right ) := \ket{\tilde \chi}.
\end{eqnarray}
with $\vert \tilde y \vert^2 = \sum_{t,s}  \left ( y_s(t) - \frac{1}{{T}} \sum_{t'}   y_s(t') \right )^2$. 
The success probability is
\begin{equation}
P_{\tilde \chi} = \frac{\delta^2 \vert \tilde y \vert^2 }{4TN}.
\end{equation}
If most of the returns are sufficiently large, $y_s(t)=\Theta(1)$, and also most $y_s(t) - \frac{1}{{T}} \sum_{t'}   y_s(t') = \Theta(1)$, then the success probability is $P_{\tilde \chi} = \Omega(1)$.
Note that $P_{\tilde \chi}$ encodes the trace of the sample covariance matrix Eq.~(\ref{eqSampleCov}) via
\begin{equation} \label{eqPtrSigma}
P_{\tilde \chi} = \frac{\delta^2 (T-1) {\rm tr} \Sigma}{4TN}.
\end{equation}
By repeating the above state preparation and postselection, this success probability can be determined and the trace ${\rm tr} \Sigma$ can be estimated. 

\bibliography{QML,Qfin}

\begin{thebibliography}{47}
\expandafter\ifx\csname natexlab\endcsname\relax\def\natexlab#1{#1}\fi
\expandafter\ifx\csname bibnamefont\endcsname\relax
  \def\bibnamefont#1{#1}\fi
\expandafter\ifx\csname bibfnamefont\endcsname\relax
  \def\bibfnamefont#1{#1}\fi
\expandafter\ifx\csname citenamefont\endcsname\relax
  \def\citenamefont#1{#1}\fi
\expandafter\ifx\csname url\endcsname\relax
  \def\url#1{\texttt{#1}}\fi
\expandafter\ifx\csname urlprefix\endcsname\relax\def\urlprefix{URL }\fi
\providecommand{\bibinfo}[2]{#2}
\providecommand{\eprint}[2][]{\url{#2}}

\bibitem[{\citenamefont{Nielsen and Chuang}(2000)}]{Nielsen2000}
\bibinfo{author}{\bibfnamefont{M.~S.} \bibnamefont{Nielsen}} \bibnamefont{and}
  \bibinfo{author}{\bibfnamefont{I.}~\bibnamefont{Chuang}},
  \emph{\bibinfo{title}{Quantum computation and quantum information}}
  (\bibinfo{publisher}{Cambridge University Press}, \bibinfo{year}{2000}).

\bibitem[{\citenamefont{Harrow et~al.}(2009)\citenamefont{Harrow, Hassidim, and
  Lloyd}}]{Harrow2009}
\bibinfo{author}{\bibfnamefont{A.~W.} \bibnamefont{Harrow}},
  \bibinfo{author}{\bibfnamefont{A.}~\bibnamefont{Hassidim}}, \bibnamefont{and}
  \bibinfo{author}{\bibfnamefont{S.}~\bibnamefont{Lloyd}},
  \bibinfo{journal}{Phys. Rev. Lett.} \textbf{\bibinfo{volume}{103}},
  \bibinfo{pages}{150502} (\bibinfo{year}{2009}).

\bibitem[{\citenamefont{Childs et~al.}(2017)\citenamefont{Childs, Kothari, and
  Somma}}]{Childs2017linear}
\bibinfo{author}{\bibfnamefont{A.}~\bibnamefont{Childs}},
  \bibinfo{author}{\bibfnamefont{R.}~\bibnamefont{Kothari}}, \bibnamefont{and}
  \bibinfo{author}{\bibfnamefont{R.}~\bibnamefont{Somma}},
  \bibinfo{journal}{SIAM Journal on Computing} \textbf{\bibinfo{volume}{46}},
  \bibinfo{pages}{1920} (\bibinfo{year}{2017}).

\bibitem[{\citenamefont{Wiebe et~al.}(2012)\citenamefont{Wiebe, Braun, and
  Lloyd}}]{Wiebe2012}
\bibinfo{author}{\bibfnamefont{N.}~\bibnamefont{Wiebe}},
  \bibinfo{author}{\bibfnamefont{D.}~\bibnamefont{Braun}}, \bibnamefont{and}
  \bibinfo{author}{\bibfnamefont{S.}~\bibnamefont{Lloyd}},
  \bibinfo{journal}{Phys. Rev. Lett.} \textbf{\bibinfo{volume}{109}},
  \bibinfo{pages}{050505} (\bibinfo{year}{2012}).

\bibitem[{\citenamefont{Rebentrost et~al.}(2014)\citenamefont{Rebentrost,
  Mohseni, and Lloyd}}]{Rebentrost2014}
\bibinfo{author}{\bibfnamefont{P.}~\bibnamefont{Rebentrost}},
  \bibinfo{author}{\bibfnamefont{M.}~\bibnamefont{Mohseni}}, \bibnamefont{and}
  \bibinfo{author}{\bibfnamefont{S.}~\bibnamefont{Lloyd}},
  \bibinfo{journal}{Physical Review Letters} \textbf{\bibinfo{volume}{113}},
  \bibinfo{pages}{130503} (\bibinfo{year}{2014}).

\bibitem[{\citenamefont{Aaronson}(2015{\natexlab{a}})}]{aaronson2015read}
\bibinfo{author}{\bibfnamefont{S.}~\bibnamefont{Aaronson}},
  \bibinfo{journal}{Nature Physics} \textbf{\bibinfo{volume}{11}},
  \bibinfo{pages}{291} (\bibinfo{year}{2015}{\natexlab{a}}).

\bibitem[{\citenamefont{Glasserman}(2003)}]{Glasserman2003}
\bibinfo{author}{\bibfnamefont{P.}~\bibnamefont{Glasserman}},
  \emph{\bibinfo{title}{Monte Carlo Methods in Financial Engineering}}
  (\bibinfo{publisher}{Springer-Verlag}, \bibinfo{year}{2003}).

\bibitem[{\citenamefont{F\"ollmer and Schied}(2004)}]{Follmer2004}
\bibinfo{author}{\bibfnamefont{H.}~\bibnamefont{F\"ollmer}} \bibnamefont{and}
  \bibinfo{author}{\bibfnamefont{A.}~\bibnamefont{Schied}},
  \emph{\bibinfo{title}{Stochastic Finance: An Introduction in Discrete Time}}
  (\bibinfo{publisher}{Walter de Gruyter}, \bibinfo{year}{2004}).

\bibitem[{\citenamefont{Hull}(2012)}]{Hull2012}
\bibinfo{author}{\bibfnamefont{J.~C.} \bibnamefont{Hull}},
  \emph{\bibinfo{title}{Options, futures, and other derivatives}}
  (\bibinfo{publisher}{Prentice Hall}, \bibinfo{year}{2012}).

\bibitem[{\citenamefont{Green}(2015)}]{Green2015}
\bibinfo{author}{\bibfnamefont{A.}~\bibnamefont{Green}},
  \emph{\bibinfo{title}{XVA: Credit, Funding and Capital Valuation
  Adjustments}} (\bibinfo{publisher}{John Wiley \& Sons},
  \bibinfo{year}{2015}).

\bibitem[{\citenamefont{Baaquie}(2004)}]{Baaquie2004}
\bibinfo{author}{\bibfnamefont{B.~E.} \bibnamefont{Baaquie}},
  \emph{\bibinfo{title}{Quantum finance}} (\bibinfo{publisher}{Cambridge
  University Press}, \bibinfo{year}{2004}).

\bibitem[{\citenamefont{Haven}(2002)}]{Haven2002}
\bibinfo{author}{\bibfnamefont{E.~E.} \bibnamefont{Haven}},
  \bibinfo{journal}{Physica A: Statistical Mechanics and its Applications}
  \textbf{\bibinfo{volume}{304}}, \bibinfo{pages}{507} (\bibinfo{year}{2002}).

\bibitem[{\citenamefont{Rebentrost
  et~al.}(2018{\natexlab{a}})\citenamefont{Rebentrost, Gupt, and
  Bromley}}]{Rebentrost2018finance}
\bibinfo{author}{\bibfnamefont{P.}~\bibnamefont{Rebentrost}},
  \bibinfo{author}{\bibfnamefont{B.}~\bibnamefont{Gupt}}, \bibnamefont{and}
  \bibinfo{author}{\bibfnamefont{T.~R.} \bibnamefont{Bromley}},
  \bibinfo{journal}{Phys. Rev. A} \textbf{\bibinfo{volume}{98}},
  \bibinfo{pages}{022321} (\bibinfo{year}{2018}{\natexlab{a}}).

\bibitem[{\citenamefont{Woerner and Egger}(2018)}]{Woerner2018}
\bibinfo{author}{\bibfnamefont{S.}~\bibnamefont{Woerner}} \bibnamefont{and}
  \bibinfo{author}{\bibfnamefont{D.~J.} \bibnamefont{Egger}},
  \bibinfo{journal}{arXiv:1806.06893}  (\bibinfo{year}{2018}).

\bibitem[{\citenamefont{Rosenberg}(2016)}]{rosenberg2016finding}
\bibinfo{author}{\bibfnamefont{G.}~\bibnamefont{Rosenberg}},
  \emph{\bibinfo{title}{Finding optimal arbitrage opportunities using a quantum
  annealer}} (\bibinfo{year}{2016}), \bibinfo{note}{1QBit white paper:
  \url{https://1qbit.com/whitepaper/arbitrage/}}.

\bibitem[{\citenamefont{Marzec}(2016)}]{Marzec2016}
\bibinfo{author}{\bibfnamefont{M.}~\bibnamefont{Marzec}},
  \emph{\bibinfo{title}{Portfolio Optimization: Applications in Quantum
  Computing}} (\bibinfo{publisher}{Wiley-Blackwell}, \bibinfo{year}{2016}),
  chap.~\bibinfo{chapter}{4}, pp. \bibinfo{pages}{73--106}.

\bibitem[{\citenamefont{Venturelli and Kondratyev}(2018)}]{Venturelli2018}
\bibinfo{author}{\bibfnamefont{D.}~\bibnamefont{Venturelli}} \bibnamefont{and}
  \bibinfo{author}{\bibfnamefont{A.}~\bibnamefont{Kondratyev}},
  \bibinfo{journal}{arXiv:1810.08584}  (\bibinfo{year}{2018}).

\bibitem[{\citenamefont{Markovitz}(1952)}]{Markowitz1952}
\bibinfo{author}{\bibfnamefont{H.}~\bibnamefont{Markovitz}},
  \bibinfo{journal}{The Journal of Finance} \textbf{\bibinfo{volume}{7}},
  \bibinfo{pages}{77} (\bibinfo{year}{1952}).

\bibitem[{\citenamefont{Wikipedia}(2018)}]{EfficientFronierWiki}
\bibinfo{author}{\bibnamefont{Wikipedia}}, \emph{\bibinfo{title}{Efficient
  Frontier}} (\bibinfo{year}{2018}),
  \bibinfo{note}{\url{https://en.wikipedia.org/wiki/Efficient_frontier }}.

\bibitem[{\citenamefont{Lloyd et~al.}(2013)\citenamefont{Lloyd, Mohseni, and
  Rebentrost}}]{Lloyd2013}
\bibinfo{author}{\bibfnamefont{S.}~\bibnamefont{Lloyd}},
  \bibinfo{author}{\bibfnamefont{M.}~\bibnamefont{Mohseni}}, \bibnamefont{and}
  \bibinfo{author}{\bibfnamefont{P.}~\bibnamefont{Rebentrost}},
  \bibinfo{journal}{arXiv:1307.0411}  (\bibinfo{year}{2013}).

\bibitem[{\citenamefont{Giovannetti
  et~al.}(2008{\natexlab{a}})\citenamefont{Giovannetti, Lloyd, and
  Maccone}}]{Giovannetti2008}
\bibinfo{author}{\bibfnamefont{V.}~\bibnamefont{Giovannetti}},
  \bibinfo{author}{\bibfnamefont{S.}~\bibnamefont{Lloyd}}, \bibnamefont{and}
  \bibinfo{author}{\bibfnamefont{L.}~\bibnamefont{Maccone}},
  \bibinfo{journal}{Phys. Rev. Lett.} \textbf{\bibinfo{volume}{100}},
  \bibinfo{pages}{160501} (\bibinfo{year}{2008}{\natexlab{a}}).

\bibitem[{\citenamefont{Giovannetti
  et~al.}(2008{\natexlab{b}})\citenamefont{Giovannetti, Lloyd, and
  Maccone}}]{Giovannetti2008_2}
\bibinfo{author}{\bibfnamefont{V.}~\bibnamefont{Giovannetti}},
  \bibinfo{author}{\bibfnamefont{S.}~\bibnamefont{Lloyd}}, \bibnamefont{and}
  \bibinfo{author}{\bibfnamefont{L.}~\bibnamefont{Maccone}},
  \bibinfo{journal}{Phys. Rev. A} \textbf{\bibinfo{volume}{78}},
  \bibinfo{pages}{052310} (\bibinfo{year}{2008}{\natexlab{b}}).

\bibitem[{\citenamefont{Martini et~al.}(2009)\citenamefont{Martini,
  Giovannetti, Lloyd, Maccone, Nagali, Sansoni, and Sciarrino}}]{Martini2009}
\bibinfo{author}{\bibfnamefont{F.~D.} \bibnamefont{Martini}},
  \bibinfo{author}{\bibfnamefont{V.}~\bibnamefont{Giovannetti}},
  \bibinfo{author}{\bibfnamefont{S.}~\bibnamefont{Lloyd}},
  \bibinfo{author}{\bibfnamefont{L.}~\bibnamefont{Maccone}},
  \bibinfo{author}{\bibfnamefont{E.}~\bibnamefont{Nagali}},
  \bibinfo{author}{\bibfnamefont{L.}~\bibnamefont{Sansoni}}, \bibnamefont{and}
  \bibinfo{author}{\bibfnamefont{F.}~\bibnamefont{Sciarrino}},
  \bibinfo{journal}{Phys. Rev. A} \textbf{\bibinfo{volume}{80}},
  \bibinfo{pages}{010302} (\bibinfo{year}{2009}).

\bibitem[{\citenamefont{Lloyd et~al.}(2014)\citenamefont{Lloyd, Mohseni, and
  Rebentrost}}]{Lloyd2014}
\bibinfo{author}{\bibfnamefont{S.}~\bibnamefont{Lloyd}},
  \bibinfo{author}{\bibfnamefont{M.}~\bibnamefont{Mohseni}}, \bibnamefont{and}
  \bibinfo{author}{\bibfnamefont{P.}~\bibnamefont{Rebentrost}},
  \bibinfo{journal}{Nature Physics} \textbf{\bibinfo{volume}{10}},
  \bibinfo{pages}{631} (\bibinfo{year}{2014}).

\bibitem[{\citenamefont{Kimmel et~al.}(2017)\citenamefont{Kimmel, Lin, Low,
  Ozols, and Yoder}}]{kimmel2017hamiltonian}
\bibinfo{author}{\bibfnamefont{S.}~\bibnamefont{Kimmel}},
  \bibinfo{author}{\bibfnamefont{C.~Y.-Y.} \bibnamefont{Lin}},
  \bibinfo{author}{\bibfnamefont{G.~H.} \bibnamefont{Low}},
  \bibinfo{author}{\bibfnamefont{M.}~\bibnamefont{Ozols}}, \bibnamefont{and}
  \bibinfo{author}{\bibfnamefont{T.~J.} \bibnamefont{Yoder}},
  \bibinfo{journal}{npj Quantum Information} \textbf{\bibinfo{volume}{3}},
  \bibinfo{pages}{13} (\bibinfo{year}{2017}).

\bibitem[{\citenamefont{Kerenidis and Prakash}(2017)}]{Kerenidis2017}
\bibinfo{author}{\bibfnamefont{I.}~\bibnamefont{Kerenidis}} \bibnamefont{and}
  \bibinfo{author}{\bibfnamefont{A.}~\bibnamefont{Prakash}}, in
  \emph{\bibinfo{booktitle}{8th Innovations in Theoretical Computer Science
  Conference (ITCS 2017)}}, edited by \bibinfo{editor}{\bibfnamefont{C.~H.}
  \bibnamefont{Papadimitriou}} (\bibinfo{publisher}{Schloss Dagstuhl},
  \bibinfo{address}{Dagstuhl, Germany}, \bibinfo{year}{2017}),
  vol.~\bibinfo{volume}{67} of \emph{\bibinfo{series}{Leibniz International
  Proceedings in Informatics (LIPIcs)}}, pp. \bibinfo{pages}{49:1--49:21}.

\bibitem[{\citenamefont{Tang}(2018{\natexlab{a}})}]{Tang2018}
\bibinfo{author}{\bibfnamefont{E.}~\bibnamefont{Tang}},
  \bibinfo{journal}{Electronic Colloquium on Computational Complexity}
  \textbf{\bibinfo{volume}{128}} (\bibinfo{year}{2018}{\natexlab{a}}).

\bibitem[{\citenamefont{Tang}(2018{\natexlab{b}})}]{Tang2018_2}
\bibinfo{author}{\bibfnamefont{E.}~\bibnamefont{Tang}},
  \bibinfo{journal}{arXiv:1811.00414}  (\bibinfo{year}{2018}{\natexlab{b}}).

\bibitem[{\citenamefont{Bai}(2003)}]{Bai2003}
\bibinfo{author}{\bibfnamefont{J.}~\bibnamefont{Bai}},
  \bibinfo{journal}{Econometrica} \textbf{\bibinfo{volume}{71}},
  \bibinfo{pages}{135} (\bibinfo{year}{2003}).

\bibitem[{\citenamefont{Grover and Rudolph}(2002)}]{Grover2002}
\bibinfo{author}{\bibfnamefont{L.}~\bibnamefont{Grover}} \bibnamefont{and}
  \bibinfo{author}{\bibfnamefont{T.}~\bibnamefont{Rudolph}},
  \bibinfo{journal}{arXiv:quant-ph/0208112}  (\bibinfo{year}{2002}).

\bibitem[{\citenamefont{Rebentrost
  et~al.}(2018{\natexlab{b}})\citenamefont{Rebentrost, Steffens, Marvian, and
  Lloyd}}]{rebentrost2018quantum}
\bibinfo{author}{\bibfnamefont{P.}~\bibnamefont{Rebentrost}},
  \bibinfo{author}{\bibfnamefont{A.}~\bibnamefont{Steffens}},
  \bibinfo{author}{\bibfnamefont{I.}~\bibnamefont{Marvian}}, \bibnamefont{and}
  \bibinfo{author}{\bibfnamefont{S.}~\bibnamefont{Lloyd}},
  \bibinfo{journal}{Physical Review A} \textbf{\bibinfo{volume}{97}},
  \bibinfo{pages}{012327} (\bibinfo{year}{2018}{\natexlab{b}}).

\bibitem[{\citenamefont{Berry and Childs}(2012)}]{Berry2012}
\bibinfo{author}{\bibfnamefont{D.~W.} \bibnamefont{Berry}} \bibnamefont{and}
  \bibinfo{author}{\bibfnamefont{A.~M.} \bibnamefont{Childs}},
  \bibinfo{journal}{Quantum Info. Comput.} \textbf{\bibinfo{volume}{12}},
  \bibinfo{pages}{29} (\bibinfo{year}{2012}), ISSN \bibinfo{issn}{1533-7146}.

\bibitem[{\citenamefont{Berry et~al.}(2015)\citenamefont{Berry, Childs, and
  Kothari}}]{Berry2015}
\bibinfo{author}{\bibfnamefont{D.~W.} \bibnamefont{Berry}},
  \bibinfo{author}{\bibfnamefont{A.~M.} \bibnamefont{Childs}},
  \bibnamefont{and} \bibinfo{author}{\bibfnamefont{R.}~\bibnamefont{Kothari}},
  \bibinfo{journal}{Foundations of Computer Science (FOCS), 2015 IEEE 56th
  Annual Symposium on} pp. \bibinfo{pages}{792--809} (\bibinfo{year}{2015}).

\bibitem[{\citenamefont{Low and Chuang}(2017)}]{Low2017}
\bibinfo{author}{\bibfnamefont{G.~H.} \bibnamefont{Low}} \bibnamefont{and}
  \bibinfo{author}{\bibfnamefont{I.~L.} \bibnamefont{Chuang}},
  \bibinfo{journal}{Physical Review Letters} \textbf{\bibinfo{volume}{118}},
  \bibinfo{pages}{010501} (\bibinfo{year}{2017}).

\bibitem[{\citenamefont{Childs}(2010)}]{Childs2010}
\bibinfo{author}{\bibfnamefont{A.}~\bibnamefont{Childs}},
  \bibinfo{journal}{Comm. Math. Phys.} \textbf{\bibinfo{volume}{294}},
  \bibinfo{pages}{581} (\bibinfo{year}{2010}).

\bibitem[{\citenamefont{Buhrman et~al.}(2001)\citenamefont{Buhrman, Cleve,
  Watrous, and de~Wolf}}]{Buhrman2001}
\bibinfo{author}{\bibfnamefont{H.}~\bibnamefont{Buhrman}},
  \bibinfo{author}{\bibfnamefont{R.}~\bibnamefont{Cleve}},
  \bibinfo{author}{\bibfnamefont{J.}~\bibnamefont{Watrous}}, \bibnamefont{and}
  \bibinfo{author}{\bibfnamefont{R.}~\bibnamefont{de~Wolf}},
  \bibinfo{journal}{Phys. Rev. Lett.} \textbf{\bibinfo{volume}{87}},
  \bibinfo{pages}{167902} (\bibinfo{year}{2001}).

\bibitem[{\citenamefont{Aaronson}(2015{\natexlab{b}})}]{Aaronson2015}
\bibinfo{author}{\bibfnamefont{S.}~\bibnamefont{Aaronson}},
  \bibinfo{journal}{Nature Physics} \textbf{\bibinfo{volume}{11}},
  \bibinfo{pages}{291} (\bibinfo{year}{2015}{\natexlab{b}}).

\bibitem[{\citenamefont{Frieze et~al.}(2004)\citenamefont{Frieze, Kannan, and
  Vempala}}]{Frieze2004}
\bibinfo{author}{\bibfnamefont{A.}~\bibnamefont{Frieze}},
  \bibinfo{author}{\bibfnamefont{R.}~\bibnamefont{Kannan}}, \bibnamefont{and}
  \bibinfo{author}{\bibfnamefont{S.}~\bibnamefont{Vempala}},
  \bibinfo{journal}{J. ACM} \textbf{\bibinfo{volume}{51}},
  \bibinfo{pages}{1025} (\bibinfo{year}{2004}).

\bibitem[{\citenamefont{Aaronson}(2009)}]{Aaronson2009}
\bibinfo{author}{\bibfnamefont{S.}~\bibnamefont{Aaronson}},
  \bibinfo{journal}{arXiv:0910.4698}  (\bibinfo{year}{2009}).

\bibitem[{\citenamefont{Bar-Yossef}(2003)}]{Yossef2003}
\bibinfo{author}{\bibfnamefont{Z.}~\bibnamefont{Bar-Yossef}}, in
  \emph{\bibinfo{booktitle}{Proc. 35th Annual ACM Symp. on Theory of
  Computing}} (\bibinfo{year}{2003}).

\bibitem[{\citenamefont{Grover}(1996)}]{Grover1996}
\bibinfo{author}{\bibfnamefont{L.~K.} \bibnamefont{Grover}}, in
  \emph{\bibinfo{booktitle}{Proceedings of the twenty-eighth annual ACM
  symposium on Theory of computing}} (\bibinfo{organization}{ACM},
  \bibinfo{year}{1996}), pp. \bibinfo{pages}{212--219}.

\bibitem[{\citenamefont{Szegedy}(2004)}]{Szegedy2004}
\bibinfo{author}{\bibfnamefont{M.}~\bibnamefont{Szegedy}}, in
  \emph{\bibinfo{booktitle}{FOCS 04 Proceedings of the 45th Annual IEEE
  Symposium on Foundations of Computer Science}} (\bibinfo{organization}{IEEE
  Computer Soc.}, \bibinfo{address}{Washington, D.C.}, \bibinfo{year}{2004}),
  pp. \bibinfo{pages}{32--41}.

\bibitem[{\citenamefont{Montanaro}(2015)}]{Montanaro2015}
\bibinfo{author}{\bibfnamefont{A.}~\bibnamefont{Montanaro}},
  \bibinfo{journal}{Proc. R. Soc. A} \textbf{\bibinfo{volume}{471}},
  \bibinfo{pages}{0301} (\bibinfo{year}{2015}).

\bibitem[{\citenamefont{Brandao and Svore}(2017)}]{Brandao2017}
\bibinfo{author}{\bibfnamefont{F.~G. S.~L.} \bibnamefont{Brandao}}
  \bibnamefont{and} \bibinfo{author}{\bibfnamefont{K.}~\bibnamefont{Svore}}, in
  \emph{\bibinfo{booktitle}{FOCS 17 Proceedings of the 45th Annual IEEE
  Symposium on Foundations of Computer Science}} (\bibinfo{organization}{IEEE
  Computer Soc.}, \bibinfo{address}{Washington, D.C.}, \bibinfo{year}{2017}).

\bibitem[{\citenamefont{van Apeldoorn et~al.}(2017)\citenamefont{van Apeldoorn,
  Gily\'en, Gribling, and de~Wolf}}]{Apeldoorn2017}
\bibinfo{author}{\bibfnamefont{J.}~\bibnamefont{van Apeldoorn}},
  \bibinfo{author}{\bibfnamefont{A.}~\bibnamefont{Gily\'en}},
  \bibinfo{author}{\bibfnamefont{S.}~\bibnamefont{Gribling}}, \bibnamefont{and}
  \bibinfo{author}{\bibfnamefont{R.}~\bibnamefont{de~Wolf}},
  \bibinfo{journal}{arXiv:1705.01843}  (\bibinfo{year}{2017}).

\bibitem[{\citenamefont{van Apeldoorn et~al.}(2018)\citenamefont{van Apeldoorn,
  Gily\'en, Gribling, and de~Wolf}}]{Apeldoorn2018}
\bibinfo{author}{\bibfnamefont{J.}~\bibnamefont{van Apeldoorn}},
  \bibinfo{author}{\bibfnamefont{A.}~\bibnamefont{Gily\'en}},
  \bibinfo{author}{\bibfnamefont{S.}~\bibnamefont{Gribling}}, \bibnamefont{and}
  \bibinfo{author}{\bibfnamefont{R.}~\bibnamefont{de~Wolf}},
  \bibinfo{journal}{arXiv:1809.00643}  (\bibinfo{year}{2018}).

\bibitem[{\citenamefont{Chakrabarti et~al.}(2018)\citenamefont{Chakrabarti,
  Childs, Li, and Wu}}]{Chakrabarti2018}
\bibinfo{author}{\bibfnamefont{S.}~\bibnamefont{Chakrabarti}},
  \bibinfo{author}{\bibfnamefont{A.~M.} \bibnamefont{Childs}},
  \bibinfo{author}{\bibfnamefont{T.}~\bibnamefont{Li}}, \bibnamefont{and}
  \bibinfo{author}{\bibfnamefont{X.}~\bibnamefont{Wu}},
  \bibinfo{journal}{arXiv:1809.01731}  (\bibinfo{year}{2018}).

\end{thebibliography}

\end{document}